\documentclass[apj]{emulateapj}
\usepackage{apjfonts}

\begin{document}

\shorttitle{GALAXY CLUSTERS, A NOVEL LOOK AT DIFFUSE BARYONS
WITHSTANDING DM GRAVITY}\shortauthors{CAVALIERE, LAPI \&
FUSCO-FEMIANO}\journalinfo{Accepted by ApJ.}

\title{Galaxy Clusters, a Novel Look at Diffuse Baryons\\Withstanding Dark Matter Gravity}
\author{A. Cavaliere\altaffilmark{1,2}, A. Lapi\altaffilmark{1,3},
R. Fusco-Femiano\altaffilmark{4}} \altaffiltext{1}{Dip. Fisica,
Univ. `Tor Vergata', Via Ricerca Scientifica 1, 00133 Roma,
Italy.} \altaffiltext{2}{Accademia Nazionale dei Lincei, Via
Lungara 10, 00165 Roma, Italy.} \altaffiltext{3}{Astrophysics
Sector, SISSA/ISAS, Via Beirut 2-4, 34014 Trieste, Italy.}
\altaffiltext{4}{INAF $-$ Istituto di Astrofisica Spaziale e
Fisica Cosmica, Via Fosso del Cavaliere, 00133 Roma, Italy.}

\begin{abstract}
In galaxy clusters the equilibria of the intracluster plasma
(ICP) and of the gravitationally dominant dark matter (DM) are
governed by the hydrostatic and the Jeans equation,
respectively; in either case gravity is withstood by the
corresponding, entropy-modulated pressure. Jeans, with the DM
`entropy' set to $K\propto r^{\alpha}$ and $\alpha\approx 1.25
- 1.3$ applying from groups to rich clusters, yields our radial
$\alpha$-\emph{profiles}; these, compared to the empirical NFW
distribution, are flatter at the center and steeper in the
outskirts as required by recent gravitational lensing data. In
the ICP, on the other hand, the entropy run $k(r)$ is mainly
shaped by shocks, as steadily set by supersonic accretion of
gas at the cluster boundary, and intermittently driven from the
center by merging events or by active galactic nuclei (AGNs);
the resulting equilibrium is described by the exact yet simple
formalism constituting our ICP \emph{Supermodel}. With a few
parameters, this accurately represents the runs of density
$n(r)$ and temperature $T(r)$ as required by up-to-date X-ray
data on surface brightness and spectroscopy for both cool core
(CC) and non cool core (NCC) clusters; the former are marked by
a middle temperature peak, whose location is predicted from
rich clusters to groups. The Supermodel inversely links the
inner runs of $n(r)$ and $T(r)$, and highlights their central
scaling with entropy $n_c\propto k_c^{-1}$ and $T_c\propto
k_c^{0.35}$, to yield radiative cooling times $t_c \approx
0.3\,(k_c/15 \, \mathrm{keV~cm}^2)^{1.2}$ Gyr. We discuss the
stability of the central values so focused: against radiative
erosion of $k_c$ in the cool dense conditions of CC clusters,
that triggers recurrent AGN activities resetting it back; or
against energy inputs from AGNs and mergers whose effects are
saturated by the hot central conditions of NCC clusters. From
the Supermodel we derive as limiting cases the classic
polytropic $\beta$-models, and the `mirror' model with $T(r)
\propto \sigma^2(r)$ suitable for NCC and CC clusters,
respectively; these limiting cases highlight how the ICP
temperature $T(r)$ strives to mirror the DM velocity dispersion
$\sigma^2(r)$ away from energy and entropy injections. Finally,
we discuss how the Supermodel connects information derived from
X-ray and gravitational lensing observations.
\end{abstract}


\section{Introduction}

Diffuse baryons and dark matter (DM) constitute the major
components of clusters and groups of galaxies, with the former
energized and shining by continual struggle against the latter.

The DM accounts for some $6/7$ of the total masses $M\sim
10^{13}- 10^{15}\, M_{\odot}$ from poor groups to rich
clusters, making for average densities $\rho \sim 10^{-26}$ g
cm$^{-3}$ with its constituent (`collisionless') particles
entertaining little or no interactions other than gravity. Thus
the DM sets the overall gravitational wells virialized within
radii $R$ up to a few Mpcs, where all bodies in dynamical
equilibrium $-$ from single particles to whole galaxies $-$
possess or acquire an $1$-D velocity dispersion $\sigma^2
\approx G\,M/5\, R \sim 10^3$ km s$^{-1}$.

The bulk of the baryons, to a fraction again close to $6/7$, is
found in the diffuse form of a hot intracluster \emph{plasma}
(the ICP), mostly comprised of protons with the neutralizing
electrons at number densities $n \sim 10^{-3}$ cm$^{-3}$, and
equilibrium temperatures $k_B T \approx m_p\, \sigma^2/2 \sim
5$ keV ($k_B$ being the Boltzmann constant and $m_p$ the proton
mass) well above most ionization potentials. This we know since
Cavaliere et al. (1971) found in the first \textsl{Uhuru} data
clear evidence of a new class of bright X-ray extragalactic
sources associated with the deep, stable gravitational wells of
the galaxy systems, emitting from the ICP they contain thermal
bremsstrahlung powers $L_X \sim 2\times 10^{-27}\, n^2 \, R^3\,
T^{1/2} \sim 10^{42}-10^{45}$ erg s$^{-1}$. The notion has been
nailed down beyond all doubts by observations of the extended
nature of these sources (Gursky et al. 1972) and of high
excitation, coronal-like lines (Mitchell et al. 1976;
Serlemitsos et al. 1977; see also Sarazin 1988) that also
pointed toward definite, somewhat subsolar metallicities.

Such temperatures and densities make the ICP an extremely
\emph{good} plasma, in fact the best in the Universe ever, as
its constituent particles in the DM gravitational wells acquire
a large kinetic relative to electrostatic energy (at mean
separations $\overline{d}=n^{-1/3}\sim 10$ cm), their ratio
being of order $k_B T/e^2\, n^{1/3}\sim 10^{12}$; this
astounding value is to be compared with its counterparts:
$10^3$ in stellar interiors, or $3\times 10^5$ in the
pre-recombination Universe. This holds despite gravity being so
exceedingly feeble at microscopic levels as to attain a mere
$G\, m^2_p/e^2\sim 8\times 10^{-37}$ of the strength that marks
the electromagnetic interactions; it holds because the plasma
condition
\begin{equation}
10^{12}\sim k_B T/e^2 n^{1/3}\equiv G\, m^2_p/e^2 \times
\bar{d}/10\,R\times \mathcal{N}~
\end{equation}
is dominated by the huge number $\mathcal{N}\equiv M/m_p\sim
10^{73}$ expressing in proton units the total DM mass with its
overwhelming gravity. As a result, the ICP at microscopic
scales constitutes a very \emph{simple}, nearly perfect gas of
particles with $3$ degrees of freedom and effective mass $\mu
m_p$ with $\mu\approx 0.6$. At intermediate scales of some $10$
kpc these sense mostly the electromagnetic interactions; by the
latter the electrons absorb or emit radiation, while the
protons share their energy over a mean free path
$\lambda_{pp}\approx 10\, (k_B T/5\,
\mathrm{keV})^2\,(n/10^{-3}\, \mathrm{cm}^{-3})^{-1}$ kpc, with
electrons following suit over some $40\, \lambda_{pp}$ toward
full thermal equilibrium.

By the same token, the macrophysics at cluster scales gets
interestingly \emph{complex}, as the ICP constitutes a faithful
\emph{archive} preserving memory of the different energy inputs
occurring throughout sizes of Mpcs and over timescales of $10$
Gyr. We shall see that the ICP physics is governed, in a
nutshell, by the interplay of electromagnetic interactions
transferring energy to the plasma, and of the DM bulk gravity
proceeding to blend and readjust it over cluster scales.

To disentangle these processes, it will prove technically
convenient to use two synthetic and formally analogous
quantities. As for the ICP, the adiabat $k \equiv k_B
T/n^{2/3}$ is straightforwardly related to the thermodynamic
specific entropy $s = 3\, k_B/2\times \log k +$ const. Usually
named `entropy' for short, this quantity is endowed with
time-honored, effective properties like: increasing
(decreasing) by energy gains (losses) other than adiabatic
compressions (expansions); and controlling the ICP settlement
into gravitational wells.

As to the DM, the analogous quantity $K=\sigma^2/\rho^{2/3}$ is
increasingly found to play $-$ in spite of all the traditional
objections leveled to defining `entropy' in a collisionless
medium dominated by long range self-gravity like the DM $-$
similar roles to a true entropy; in fact, it increases during
the fast collapse with associated major mergers that set up the
halo bulk, and stays put during the subsequent slow mass
accretion producing a quiet development of the outskirts from
the inside out. These two stages have been recently recognized
in sophisticated $N$-body simulations that follow the
development of halos embedding galaxies or galaxy systems from
initial cold DM perturbations (e.g., Zhao et al. 2003; Diemand
et al. 2007). From simulations, during quiet stages the simple
powerlaw run $K(r)\propto r^{\alpha}$ is found to apply
remarkably well with definite slopes close to $\alpha \approx
1.25$ throughout the structure's main body, as first stressed
by Taylor \& Navarro (2001) and confirmed by many others (e.g.,
Dehnen \& McLaughlin 2005; Hoffman et al. 2007; Ascasibar \&
Gottl\"{o}ber 2008; Vass et al. 2009; Navarro et al. 2009).

We shall find these two entropies to be most useful in
computing and relating the macroscopic, static equilibria of DM
and baryons in the \emph{same} gravitational wells as set by
the former. This is perceived on just inspecting their related
equilibrium conditions for the density runs $n(r)$ and
$\rho(r)$, namely, the hydrostatic vs. Jeans equation in the
form
\begin{equation}
{1\over \mu m_p \, n}\, {\mathrm{d}\over \mathrm{d}r}
(n^{5/3}\, k) = - {G\, M(<r)\over r^2} = {1\over \rho}
{\mathrm{d}\over \mathrm{d}r} (\rho^{5/3}\,K) ~.
\end{equation}
While the rich contents of the second equality have been
discussed in detail by Lapi \& Cavaliere (2009a [hereafter
LC09], see also \S~2 for a recap), in the present paper we will
focus on the first equation. To this purpose, we make use of a
physical model for $k(r)$ to derive the detailed pressure
structure of the ICP.

Our approach develops the line extensively pursued in the
literature at increasingly sophisticated levels. A simple
isothermal or polytropic state has been adopted by, e.g.,
Cavaliere \& Fusco-Femiano (1976, 1978), Balogh et al. (1999),
Dos Santos \& Dor\'e (2002), and Ostriker et al. (2005).
Refined models were based on a more articulated entropy run,
consisting of an outer ramp produced by \emph{gravitational}
accretion shocks going on in the aftermath of cluster formation
(see Tozzi \& Norman 2001, Voit et al. 2003, Lapi et al. 2005
[hereafter LCM05]), plus a central entropy floor of
\emph{nongravitational} origin (e.g., Babul et al. 2002; Voit
et al. 2002; Voit 2005). The latter may be contributed by a
number of processes: cooling that selectively removes some
low-entropy gas from the inner regions (see Bryan 2000; Voit \&
Bryan 2001; McCarthy et al. 2004); preheating of the ICP due to
supernovae (SNe)/active galactic nuclei (AGNs) within the
cluster progenitors (see Cavaliere et al. 1997; LCM05; McCarthy
et al. 2008); the direct action within the cluster of central
AGNs (Valageas \& Silk 1999; Wu et al. 2000; Scannapieco \& Oh
2004; LCM05). Additional gravitational events in the form of
major mergers (Balogh et al. 2007; McCarthy et al. 2007) that
now and then punctuate the equilibrium state of formed clusters
(Cavaliere et al. 1999) may reach down to the center and
enhance the entropy there.

Here we improve over these previous studies in the following
respects. As to the DM potential well, we adopt the physical
$\alpha$-profiles obtained from solving the Jeans equation
rather than assuming the empirical NFW fit (see \S~2). As to
the ICP equilibrium, we present a novel analytical solution for
the ICP density and temperature runs in terms of the ICP
entropy distribution (see \S~3). The distribution that we use
comprises an outer ramp and a central floor, marked by two
parameters physically assessed in terms of steady, self-similar
accretion shocks plus additional energy inputs due to central
AGNs and mergers (see \S~3.1). We insert the entropy
distribution in our equilibrium solution to obtain the ICP
`Supermodel', that effectively represents with two ICP
parameters the extended profiles of the X-ray surface
brightness and temperature for either classes: the cool core
(CC) and the non cool core (NCC) clusters, see \S~4. We discuss
the stability against cooling and feedback of the ensuing
central conditions (see \S~5). We recover classic models for
the ICP distribution of density and temperature as limiting
cases of the Supermodel valid in different radial ranges and
for different central entropy levels (see \S~6). We provide
specific predictions for future X-ray observations of the ICP,
and model-independent tests of the entropy runs underlying the
ICP Supermodel (see \S~7).

Throughout this work we adopt a standard, flat cosmology (see
Spergel et al. 2007) with normalized matter density $\Omega_M =
0.27$, dark energy density $\Omega_\Lambda = 0.73$, and Hubble
constant $H_0 = 72$ km s$^{-1}$ Mpc$^{-1}$.

\begin{figure}
\epsscale{1.1}\plotone{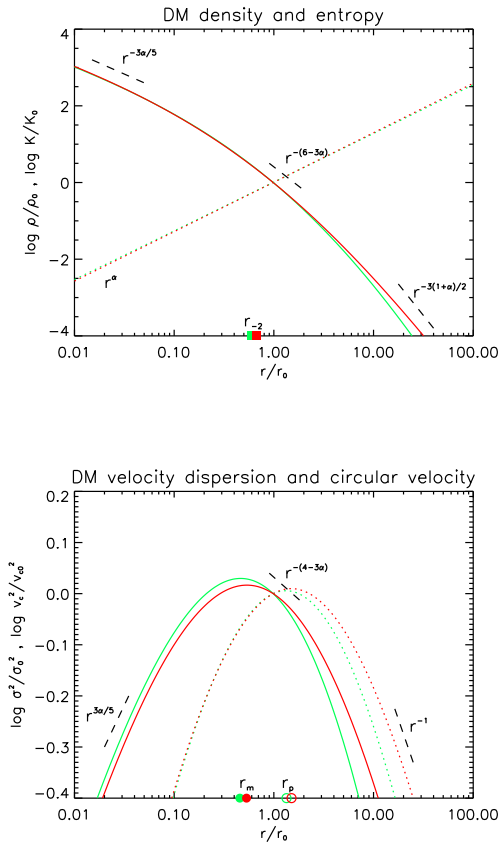}\caption{Top panel: radial runs
of the DM density (solid) and entropy (dotted); bottom panel:
radial runs of the DM velocity dispersion (solid) and circular
velocity (dotted). Both panels refer to the $\alpha$-profiles
with $\alpha=1.27$ (green lines) and $\alpha=1.3$ (red
lines); the profiles are normalized to unity at the radius
$r_0$ where the density slope $\gamma_0$ applies. The
behaviors in the inner regions, main body and outskirts are
highlighted by black dashed lines. The radii $r_{-2}$ where the
density slopes equal $-2$ (used in defining the concentration
$c$, see \S~2), $r_m$ where the velocity dispersion peak, and
$r_p$ where the circular velocity peak are highlighted as
squares, filled and empty circles, respectively.}
\end{figure}

\section{{\large $\alpha$-}profiles for the DM halos}

From LC09 we recall for use in the present paper a number of
basic features in the equilibrium of the DM halos embedding the
ICP, as described by the Jeans equation with entropy $K \propto
r^{\alpha}$ (see \S~1); in compact form this reads
\begin{equation}
\gamma = {3\over 5}\, \alpha + {3\over 5}\, {v^2_c\over \,
\sigma^2} ~.
\end{equation}
With $\alpha\equiv \mathrm{d}\log K/ \mathrm{d}\log r$ set to a
constant, this yields the changing density slope
$\gamma(r)\equiv - \mathrm{d}\log \rho/ \mathrm{d}\log r$ in
terms of the increasing ratio to $\sigma^2 (r)$ of the halo
circular velocity squared $v^2_c (r)= G\, M(<r)/r$ that
provides normalization and running estimates for the
gravitational potential. The above equation provides the
detailed cluster gravitational well, within which the ICP is to
adjust.

The \emph{physically} relevant values of $\alpha$ are pinned
down from the structures' cosmogonic development. Based on the
straightforward scaling laws $\sigma^2\propto M/r$ and $r
\propto M/\dot{M}^{2/3}$, it is seen that the entropy is to
scale as $K\propto r\, M^{1/3}$, a run clearly implying a
radial slope $\alpha\ga 1$. The detailed time dependence of the
accretion rate $\dot{M}/M$ and the related quantities may be
derived semianalytically within the standard $\Lambda$CDM
cosmogony; the result for the allowed range is $\alpha = 1.25 -
1.3$ from groups to massive clusters, narrowed down to
\begin{equation}
\alpha \approx 1.27 - 1.3
\end{equation}
for average masses $M\sim 10^{14}-10^{15}\,M_{\odot}$ from poor
to rich clusters of main interest here. In the structure's
development these values obtain at the transition epoch from
the stage of fast collapse to that of slow accretion (relative
to the running Hubble time), when the potential well attains
its maximal depth marked by $\dot{M}\, t/M\sim 1$. Such a
two-stage development turns out to be in tune with the
intensive, detailed $N$-body simulations recalled in \S~1. In
addition, the simulations support a value of $\alpha$ in the
range above, which stays put from the center throughout the
halo bulk as the late quasi-equilibrium configuration develops
from the inside out.

These values may be inserted into the Jeans equation, to find
(as pioneered by Taylor \& Navarro 2001 and Dehnen \&
McLaughlin 2005) all the \emph{viable} solutions for $\rho(r)$,
that we name '$\alpha$-profiles' and illustrate in Fig.~1.
These feature a monotonic radial run satisfying physical
central and outer boundary conditions, i.e., zero gravitational
force (corresponding to a flat minimum of the potential) and
finite (hence definite) overall mass, respectively.

The ensuing behavior of the $\alpha$-profiles, basic to the ICP
equilibrium, is illustrated in the top panel of Fig.~1, and
highlighted by the analytic expressions of the slopes
\begin{equation}
\gamma_a = {3\over 5}\,\alpha~,~~~~\gamma_0 =6-3\alpha~,~~~~
\gamma_b={3\over2}\,(1+\alpha)~.
\end{equation}
These start from the central ($r\rightarrow 0$) value
$\gamma_a\approx 0.76 - 0.78$, progressively steepen to
$\gamma_0\approx 2.19 - 2.1$ at the point $r_0$ that marks the
halo main body, and steepen further into the outskirts to the
value $\gamma_b\approx 3.41 - 3.44$ at around the virial radius
$R$ before going into a final cutoff. The $\alpha$-profiles are
seen (see LC09) to correspond to a maximal value
$\kappa_{\mathrm{\rm crit}}(\alpha) = v^2_c/\sigma^2\approx
2.6-2.5$ for the relative gravitational pull at the point $r =
r_p\ga r_0$ where $v_c^2(r)$ peaks (see also bottom panel of
Fig.~1). After Eq.~(3) this also implies at $r_p$ a maximal
slope $\gamma_p = 3\,(\alpha+k_{\rm crit})/5\approx 2.32 -
2.28$.

At variance with the empirical NFW formula (Navarro et al.
1997) that features angled central potential or pressure and
diverging mass,for the $\alpha$-profiles the inner slope is
considerably \emph{flatter}, and the outer one is
\emph{steeper} as to result in a definite overall mass. The
radial range $r>r_{-2}$ where the density profile is steeper
than a reference slope $\gamma=2$ may be specified in terms of
the usual concentration parameter $c\equiv R/r_{-2}$, that may
be viewed as a measure of central condensation and/or
outskirts' extension. In up-to-date numerical simulations (see
Zhao et al. 2003; Diemand et al. 2007) this is found to take on
value $c\approx 3.5$ at the transition redshift $z_t$, and to
increase thereafter to current values $c\approx 3.5\,(1+z_t)$
up to $c\approx 10$ for the fraction about $10\%$ of rich
clusters with early transition epoch $z_t\sim 1.5$. Instead,
values $c\approx 4-5$ apply to the much more numerous clusters
with recent transition epochs.

Another basic feature of the $\alpha$-profiles is provided by
the \emph{peaked} run of the gravitationally acquired
dispersion
\begin{equation}
\sigma^2(r)\propto K(r)\, \rho^{2/3}(r)~,
\end{equation}
see bottom panel of Fig.~1. Towering above the central and the
outer drops related to the cold nature of the DM, the peak
results from the central rise of $K(r)$ and the outer steep
falloff of $\rho(r)$, and will have a striking counterpart in
the ICP. The characteristic values of $\sigma^2$ for the
$\alpha$-profiles are set in terms of \emph{minimal} values for
$\sigma^2=v_c^2/\kappa_{\rm crit}$; this is related to limited
randomization during accretion of the infall kinetic energy
gained by the initially cold DM particles.

These $\alpha$-profiles actually depend weakly on $\alpha$ in
the named range (see Fig.~1), and though derived from the
isotropic Jeans Eq.~(3) prove to be \emph{stable} against
addition of reasonable anisotropies. The latter are described
by the standard Binney (1978) parameter $\hat{\beta}$, which
numerical simulations suggest to increase outwards from central
values $\hat{\beta}(0) \ga -0.1$ meaning weakly tangential
anisotropy, toward $\hat{\beta}\la 0.5$ that implies prevailing
radial motions (see Dehnen \& McLaughlin 2005; Hansen \& Moore
2006; also H{\o}st et al. 2009). Such density profiles turn out
to be slightly \emph{flattened} at the center and considerably
\emph{steepened} into the outskirts.

The $\alpha$-profiles are also \emph{stable} against educated
variations in the condition $\alpha =$ const. Actually, the
cosmogonic buildup given by our semianalytic computation yields
a slow decrease of $\alpha(r)$ into the outskirts, as these
develop from the inside out at late times after the transition.
Again in keeping with the simulations, the outer development
does not affect the inner gravitational potential and the
equilibrium described by the Jeans equation, that (with its
central boundary condition) also works from the inside out. In
fact, for the present use we have computed the density profiles
associated with $\alpha(r)$, and checked these to be very close
to the parent $\alpha$-profiles but for being somewhat steeper
into the outskirts, with densities lower by about $15\%$ in the
vicinity of the virial radius.

It is to be stressed that the $\alpha$-profiles for $M\sim
10^{14}-10^{15}\, M_{\odot}$, especially with full $\alpha(r)$
and anisotropies, turn out to provide an optimal fit to the
surface density runs as derived from gravitational lensing
observations in around massive clusters that just require
central slopes flatter and outer slopes steeper than the
standard NFW (cf. Broadhurst at al. 2008). Such a fit for A1689
requires high values $c\sim 10$ and early transition epochs
$z_t\approx 1.5$, related conditions that the two-stage
development predicts to occur in about $10\%$ of the rich
clusters (Lapi \& Cavaliere 2009b).

We will adopt these $\alpha$-profiles depending on the
\emph{two} DM key parameters $\alpha$ and $c$ to describe the
gravitational potential containing the ICP, to which we now
turn.

\section{The ICP equilibrium}

We will focus on clusters in equilibrium conditions, in between
the punctuating violent mergers for which we refer the reader
to the review by Markevitch \& Vikhlinin (2007). When the DM
halo is close to equilibrium, even more so will be the
pervading ICP; in fact, the sound crossing time $R/(5\, k_B T/3
\,\mu m_p)^{1/2} $ is seen to be somewhat shorter than the
dynamical time $R/\sigma$ on recalling from \S~1 that $k_B T
\approx \mu m_p\, \sigma^2$ holds. In these conditions the ICP
is governed by the hydrostatic, Jeans-like equation provided by
the first and the second member of Eq.~(2). In a compact form
similar to Eq.~(3) this writes
\begin{equation}
g = {3\over 5}\, a + {3\over 5}\,b~,
\end{equation}
once the ICP entropy run $k(r)$ is given and used to factor out
the product $\mathrm{d}\, (k\, n^{5/3})/\mathrm{d}r $ in the
original Eq.~(2). In fact, the entropy slope $a \equiv
\mathrm{d} \log k/ \mathrm{d}\log r$ enters Eq.~(7) to yield
the density slope $g(r)\equiv - \mathrm{d}\log n /
\mathrm{d}\log r$, concurring with the gravitational pull
measured by the (squared) ratio $ 3\, b(r)/5 \equiv 3\, \mu
m_p\, v^2_c(r)/ 5 \, k_B T(r)$ of the DM circular velocity
(discussed in \S~2) to the sound speed.

The above constitutes just a first order differential equation,
that transforms to linear in terms of the variable
$n^{2/3}(r)$. By textbook recipes (cf. Dwight 1961) this is
amenable to a simple solution in form of a quadrature, that is
conveniently written as
\begin{equation}
\bar{T}(\bar{r}) = \bar{k}(\bar{r})\, \bar{n}^{2/3}(\bar{r}) =
\bar{k}^{\,3/5}(\bar{r}) \, [1 + {2\over 5}\, b_R\,
\int^1_{\bar{r}}\, {\mathrm{d}\bar{r}'\over \bar{r}'}\,
\bar{v}^2_c(\bar{r}')\, \bar{k}^{-3/5}(\bar{r}')]~.
\end{equation}
Here barred variables are normalized to their boundary value at
$r= R$ where $b(r)$ takes on the value $b_R$, while $v^2_c(r)$
is taken from the $\alpha $-profiles with its weak dependence
on $\alpha$.

\begin{figure}
\epsscale{1.}\plotone{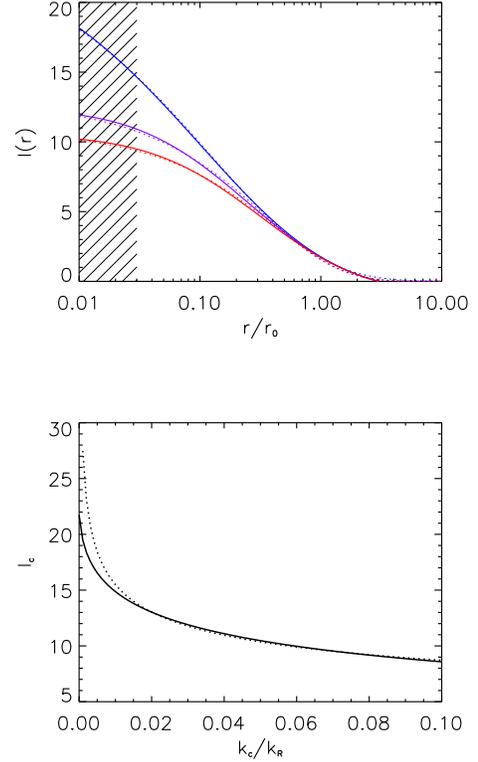}\caption{Top panel: the integral
entering Eq.~(8) is plotted as a function of $r$ for the values
$\bar{k}_c = 5\times 10^{-2}$ (red), $2.5\times 10^{-2}$
(cyan), and $0$ (blue) used in the next Figures; the dotted lines
represent the analytic fit provided in Appendix A. Bottom
panel: The value of the integral at the center is plotted
as a function of $\bar{k}_c$ (solid line), and is compared with
the powerlaw approximation $\bar{k}_c^{-1/4}$ (dotted line);
the latter holds to better than $5\%$ in the most relevant
range $\bar{k}_c\approx 0.01-0.15$, see \S~4.}
\end{figure}

In Fig.~2 (top panel) we plot the integral appearing in the
above equation, along with the simple analytical fit presented
in Appendix A. Note three circumstances. First, $2\, b_R/5
\approx 1$ holds (e.g., $2\, b_R/5 \approx 1.06 - 1.02$ on
using $b_R= 2.65 - 2.55$, see Eq.~[10]). Second, at the center
the integral dominates in the square bracket; thus the memory
of the boundary condition is swamped while the integral is
numerically found to scale as $k_c^{-1/4}$ with the central
entropy $k_c$, see Fig.~2 (bottom panel). Third, on approaching
the center where the $r$-dependence of the bracket is already
saturating, Eq.~(8) implies $T(r)\propto n^{-1}(r)\propto
k^{3/5}(r)$. The issue to stress is that, rather than a model,
the above Eq.~(8) has the standing of a \textit{theorem} in
hydrostatics valid for the ICP of clusters close to
equilibrium.

\subsection{ICP entropy}

What actually needs \textit{physical} modeling is the entropy
run $k(r)$. We base upon the notions that entropy is erased by
radiative cooling on the timescale $t_c\approx 65\, (k_B T/5\,
\mathrm{keV})^{1/2}\,$ $(n/10^{-3}\, \mathrm{cm}^{-3})^{-1}$
Gyr (Sarazin 1988), while substantial raising requires
\emph{shocks} (see Appendix B) as are driven by supersonic
outflows from center and set by inflows across the boundary.

\subsubsection{Entropy deposited at the boundary}

At the outer end, the mathematics of Eqs.~(7) requires one
boundary condition fixing $b_R$. On the observational side,
$T(r)$ is found to decline slowly toward the virial radius
(Molendi \& Pizzolato 2001), and at such a rate it would take
tens of Mpcs to decline smoothly from keV values to those some
$10^{-2}$ times lower as prevailing in the external medium. So
at $r\approx R$ a discontinuity is to occur and terminate the
hot ICP, at variance with the smooth if steeper decline of the
DM density. In fact, most of the transition takes place across
a few mean free paths $\lambda_{pp}$ at the \textit{accretion
shocks} produced when external gas supersonically falls into
the DM potential.

Numerical simulations (e.g., Tormen et al. 2004; see their Fig.
8) show that accretion of the smooth gas and minor lumps making
up a major fraction of the accreted mass drives a complex
patchwork of shocks mostly comprised within an outer layer with
thickness of order $10^{-1}\,R$; across such a layer the
standard conservation laws of mass, momentum and energy may be
applied leading to the classic Rankine-Hugoniot jump
conditions; to within $10\%$ the results are similar to a
coherent accretion shock, roughly spherical and located at
$r\sim R$ (see LCM05 for a complete treatment, and Ettori \&
Fabian 1998 for possibly delayed electron equilibrium). The
impact of major lumps reaching down to the center will be dealt
with in \S~3.1.3.

The jump conditions take on a particularly simple form for cold
inflow into rich clusters that cause \emph{strong} shocks with
Mach numbers squared $\mathcal{M}^2$ considerably exceeding
unity (see Appendix B). In strong shocks lingering at $r
\approx R$ (and expanding along with the cluster's
development), maximal conversion of infall energy occurs over a
radial range of order $\lambda_{pp}$, to yield
\begin{equation}
k_B T_R = {2\over 3}\, \mu m_p\,v_R^2\,\Delta\phi_{\rm tR}~,
\end{equation}
where $2\,v_R^2\,\Delta\phi_{\rm tR}$ is the kinetic energy per
unit mass of the gas freely falling across the potential drop
down to $r = R$ from the turning point where infall starts (see
LCM05). Thus we find
\begin{equation}
b_R = {3\over 2\, \Delta \phi_{\rm tR}}~,
\end{equation}
with values $2.65-2.55$ corresponding to $\alpha = 1.27-1.3$
and $\Delta\phi_{\rm tR}\approx 0.57 - 0.59$.

At the boundary the equilibrium Eq.~(7) yields $g = 3\,(a +
b_R)/5$. Meanwhile, powerlaw approximations $k \propto r^a$, $n
\propto r^{-g}$ apply in the vicinity of $r=R$, so $m(r)\propto
r^{3-g}$ describes the ICP mass $m$ in the outer layer, and
correspondingly
\begin{equation}
k\propto m^{a/(3-g)}\propto m^{5 a/3 (5-a-b_R)}
\end{equation}
obtains.

On the other hand, pursuing the scaling for $K(r)$ recalled in
\S~2, LC09 show that the DM entropy behaves as $K \propto
M^{4/3}/ \dot M^{2/3} \propto M^{3/2}$ on considering that
$M\propto t^{4/5}$ holds in the standard $\Lambda$CDM cosmogony
for $z\la 0.5$ during the slow accretion stage. As external gas
and DM are accreted in cosmic proportion, in the outer layer
$\dot{m} \propto \dot{M}$ applies, and it follows that the ICP
entropy at the boundary can be expressed as
\begin{equation}
k \propto m^{3/2} ~.
\end{equation}
Then on equating the exponents in Eqs.~(11) and (12) one finds
\begin{equation}
a_R = {45-9\,b_R\over 19}~~~~~\mathrm{and}~~~~~~g_R =
{27+6\,b_R\over 19}~.
\end{equation}
With the values of $b_R$ discussed above, these yield the
slopes $a_R\approx 1.1$ and $g_R\approx 2.2$, both considerably
\emph{flatter} than the corresponding DM values. Note from
Eq.~(13a) how the slope $a_R$ at the boundary decreases if
$b_R$ is increased. We expect such a decrease to take place for
clusters with high concentrations corresponding to shallow
outer potentials (see Lapi \& Cavaliere 2009b), that decrease
the outer drops $\Delta\phi_{\rm tR}$ entering $b_R$ through
Eq.~(10). E.g., a value $c\approx 10$ (holding for a cluster
with an early transition, see \S~2) in place of the usual
$c\approx 4$ (holding for clusters with a recent transition)
implies $\Delta\phi_{\rm tR}$ to lower to $0.47$, $b_R$ to grow
to $3.2$, and $a_R$ to decrease to $0.85$ as may be the case
for A1689 (see Lemze et al. 2008; Lapi \& Cavaliere 2009b);
meanwhile, from Eq.~(13b) $g_R$ increases to $2.4$. A flat
entropy slope may be also produced when the boundary shock is
weakened by substantial preheating of the infalling gas; such a
condition is relevant for poor clusters and groups with
comparatively low potentials (see LCM05), e.g., $a\approx 0.8$
applies to systems with $M\la 10^{14}\, M_{\odot}$ with
external preheating at levels around $1/2$ keV per particle.

On the other hand, after Eqs.~(13) high values of $a$ with an
upper bound at $45/19\approx 2.4$ (in keeping with the variance
observed by Cavagnolo et al. 2009) correspond to low values of
$b_R$, hence to large $\Delta\phi_{\rm tR}$; these imply flat
densities and temperatures sustained to high values in the
outskirts.

\subsubsection{Entropy stratification}

No other major sources or sinks of energy and entropy occur
inward of the boundary at a few Mpcs down to the central $10^2$
kpc. So throughout the ICP bulk, the entropy slope is to stay
at its boundary value $a = a_R$, and in rich clusters with
standard concentration
\begin{equation}
k (r) \propto r^{1.1} ~
\end{equation}
is to hold, a result similar to Tozzi \& Norman (2001), LCM05,
and Voit (2005).

On the other hand, while the simple thermodynamics of the ICP
maintains the run of $k(r)$ in this powerlaw form, its more
complex adiabatic readjustments within the gravitational well
cause the density slope $g(r)$ to flatten out inward; this is
granted by Eq.~(7), as the inequality $b(r) < b_R$ (i.e.,
$v^2_c(r)/k_B T (r) < 1$) progressively strengthens away from
the boundary. As a result, the ICP slope $g(r)$ not only starts
out but also stays generally \emph{flatter} than DM's
$\gamma(r)$ inward of the boundary, as given by
\begin{equation}
g(r) =3 \, [a + b(r)]/5 < 3\, [\alpha + \kappa (r)]/5 = \gamma(r)~.
\end{equation}
This implies the ICP density run $n(r)$ at the boundary to
parallel the DM's $\rho (r)$ at some other point well inside
the cluster's main body.

\begin{figure}
\epsscale{1.2}\plotone{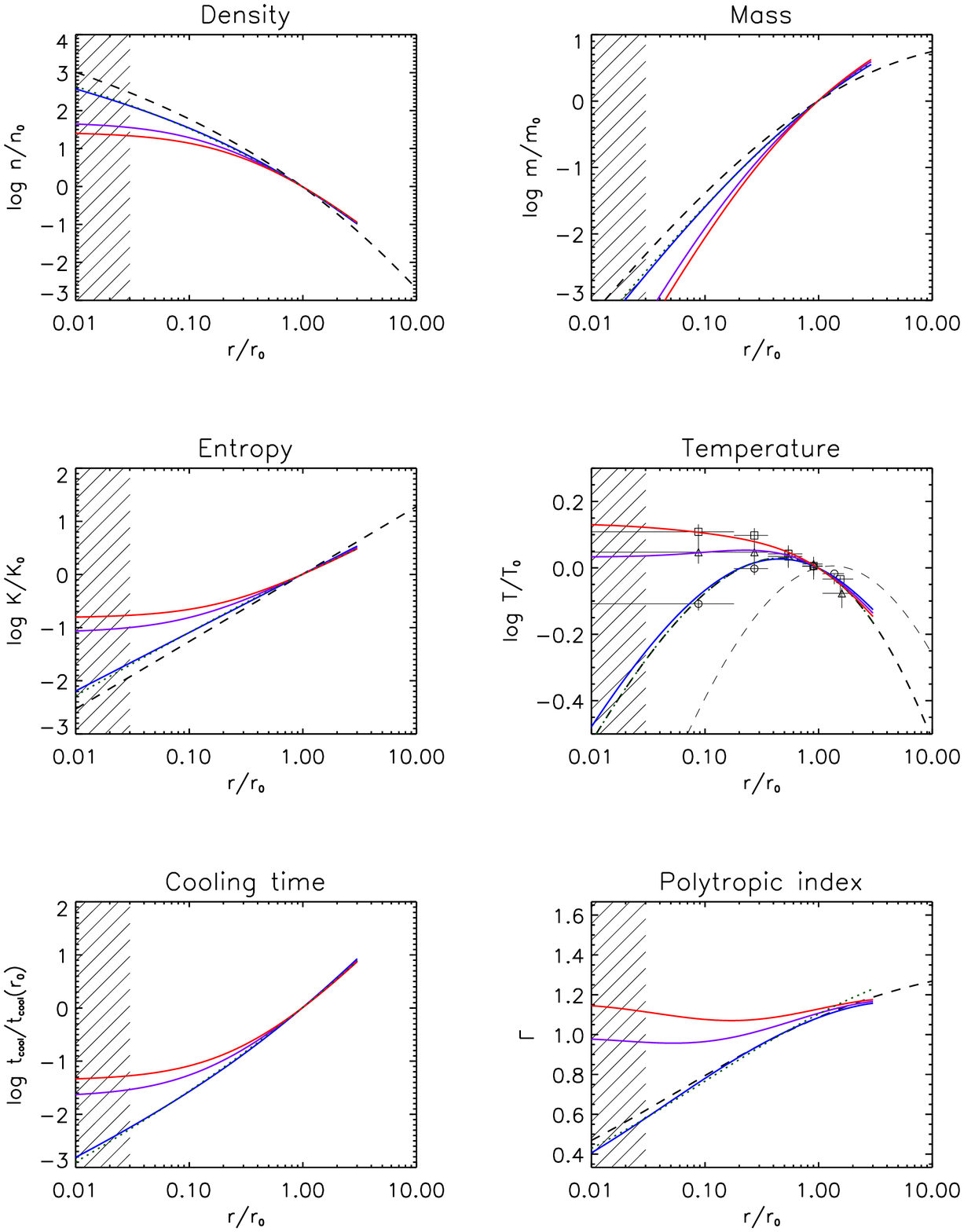}\caption{Radial
runs of the ICP density, mass, entropy, temperature, cooling
time, and polytropic index given by the Supermodel. Solid lines
are for different values of the central entropy
$\bar{k}_c=5\times 10^{-2}$ (red), $2.5\times
10^{-2}$ (cyan), and $0$ (blue); dotted lines refer to
the `mirror' model discussed in \S~6.1 with $\beta=0.75$;
dashed lines illustrate the underlying DM distribution taken
from the $\alpha$-profile with $\alpha=1.27$ and an average
concentration value $c\equiv R/r_{-2} = 5$ (see \S~2). In the panel
representing $T(r)$, we plot the DM velocity dispersion
$\sigma^2(r)$ as a thick dashed line, and the circular
velocity $v_c^2(r)$ as a thin dashed line; we also show average
data for the CC (circles), NCC (squares), and UNC (triangles)
classes from Leccardi \& Molendi (2008). The hatched area
outlines the central range hardly accessible to current
resolutions.}
\end{figure}

\subsubsection{Entropy in the central region}

Next we discuss the central value $k_c$ of the entropy and its
origins. Recall from \S~1 that the central entropy may be
produced by shocks driven by substantial merging events
reaching the center (McCarthy et al. 2007; Balogh et al. 2007),
and by AGNs residing in the central bulge-dominated galaxies
(see Valageas \& Silk 1999; Wu et al. 2000; Cavaliere et al.
2002; Scannapieco \& Oh 2004; LCM05). In addition, a basal
entropy level may be left over by SNe and AGNs that preheated
the gas in the volume due to collapse into, or to accrete onto
the cluster (Cavaliere et al. 1997, LCM05, McCarthy et al.
2008).

To begin with, consider preheating; this may be expressed in
terms of entropy advected across the shock by the currently
infalling gas (see Appendix B). However, such a process is
known to be effective only within limits: if always strong, it
would flatten the bulk entropy slope of most clusters to values
$a<1$ (see \S~3.1.1 and Fig.~3), and produce really flat
entropy profiles in most groups at variance with the
observations (see Balogh et al. 1999; Pratt \& Arnaud 2003;
Rasmussen \& Ponman 2004); unless tailored to cluster
precollapse size, it would involve larger volumes and require
high total energetics; finally, the presence of Ly$\alpha$
absorbers limits actual preheating to optimal redshifts $z\sim
2-3$ when it may catch the peak of star/AGN formation while
still operating at moderate cosmological densities (see Voit
2005).

\begin{figure}
\epsscale{1.2}\plotone{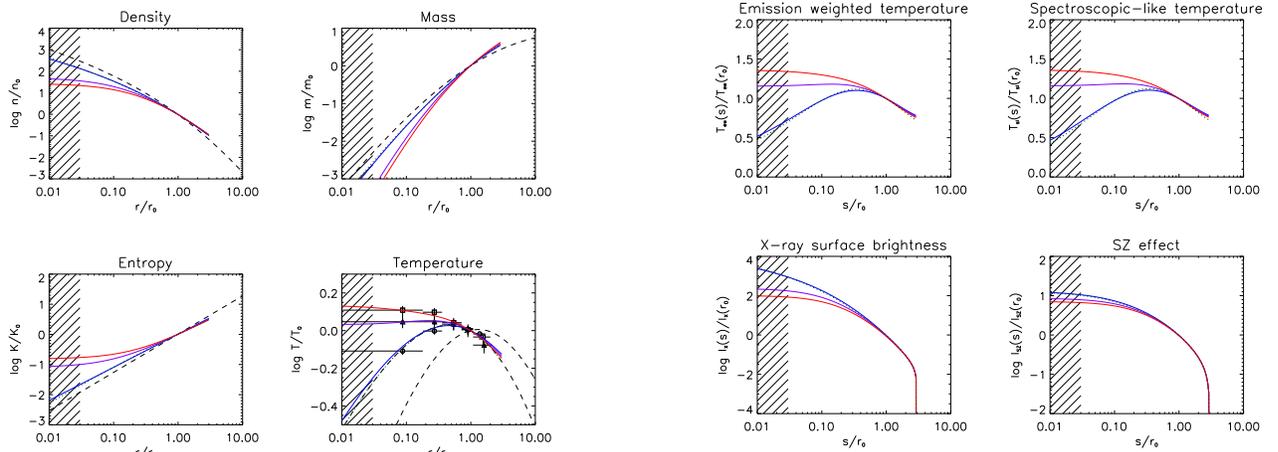}\caption{ICP projected
distributions of emission-weighted temperature,
spectroscopic-like temperature, X-ray surface brightness, and
l.o.s. SZ effect. Linestyles as in Fig.~3.}
\end{figure}

Consider then the shocks produced at cluster centers by major
mergers. A main progenitor with mass already close to the
present value $M$ undergoes only a few more, if any merging
shocks; each of these, in view of the considerable mass import
and high infall velocity, can deliver large energies up to
$10^{64}$ ergs or several keV per particle, and may produce
levels of $k_c$ up to several $10^2$ keV cm$^2$ (see Appendix
B). With timescales for free radiative cooling of order
\begin{equation}
t_{c}\approx 0.3\, \left({k_c \over 15\,
\mathrm{keV~cm}^2}\right)^{3/2}\, \left({k_B T\over 5\,
\mathrm{keV}}\right)^{-1}\, \mathrm{Gyr} ~,
\end{equation}
(Voit \& Donahue 2005), the above levels take several Gyrs to
decrease substantially. At the other end, a much lower entropy
state would rapidly run down to values lower still. So a key
point is to secure intermediate but persisting entropy levels.

Values of $k_c$ in the range $10 - 30$ keV cm$^2$ as frequently
observed despite fast radiative erasure point toward entropy
inputs frequently refreshed; a promising path is provided by
sizeable and recurrent energy injections by central AGNs,
triggered into activity by renewed fueling in a low-entropy,
dense environment of their underlying supermassive black holes
(BHs). These are observed to undergo outbursts able to inject,
nearly independent of cluster mass, energies $\Delta E$ up to
$10^{62}$ erg or a few keV per particle, yielding $k_c$ of
several tens keV cm$^2$. Cooling on the scale of $10^{-1}$ Gyr
may be offset as the several supermassive BHs inhabiting the
many bulge-dominated galaxies in the central region of a rich
cluster alternatively kindle up, and collectively recur over
such timescales (see Nusser et al. 2006; Ciotti \& Ostriker
2007; Conway \& Ostriker 2008). In particular, outbursts
lasting for a few $10^8$ yr can continuously drive out to $300$
kpc pressurized blastwaves terminating into a shock with Mach
numbers sustained at $\mathcal{M}^2 \ga 3$, as computed in
detail by LCM05 and observed by Forman et al. (2005), Nulsen et
al. (2005), McNamara \& Nulsen (2007), see also Appendix B.

All these processes contribute to raise the entropy over an
extended region around the center, adding to the outer run
given by Eq.~(14); the combined entropy profile may be
described by the simple parametric expression (see Voit 2005
and references therein)
\begin{equation}
\bar{k}(\bar{r}) = \bar{k}_c + (1-\bar{k}_c)\, \bar{r}^a ~
\end{equation}
represented in Fig.~3. Having computed the outer powerlaw slope
$a$ around $1.1$ (see \S~3.1.1) and having discussed central
entropy levels $k_c$ from a few tens to several hundreds keV
cm$^2$, we next use these \emph{two} ICP key parameters in our
equilibrium `Supermodel'.

\begin{figure}
\epsscale{1.2}\plotone{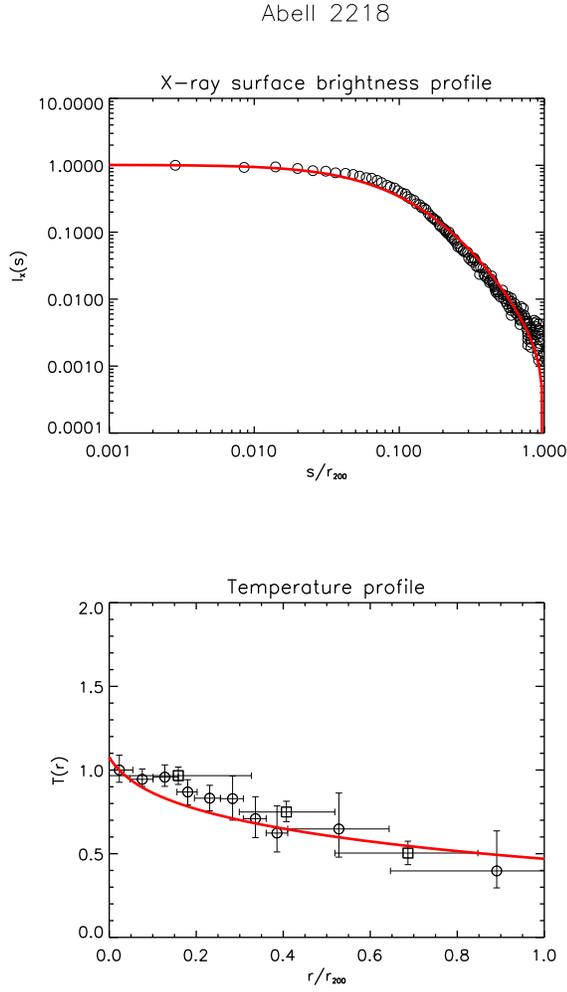}\caption{X-ray
surface brightness and temperature of the NCC cluster A2218.
Circles represent \textsl{XMM-Newton} data (Zhang et al. 2008),
while squares refer to preliminary \textsl{Suzaku}
observations. For the sake of clarity, in the top panel we do
not report the formal error bars; these are generally small,
except for the outermost data limited by sensitivity, and for
the innermost ones limited by resolution. The solid
line represents the Supermodel for a given DM
$\alpha=1.27$-profile with concentration $c= 3.5$, and ICP
entropy with slope $a= 1.1$ and central value $\bar{k}_c=
10^{-1}$.}
\end{figure}

\begin{figure}
\epsscale{1.2}\plotone{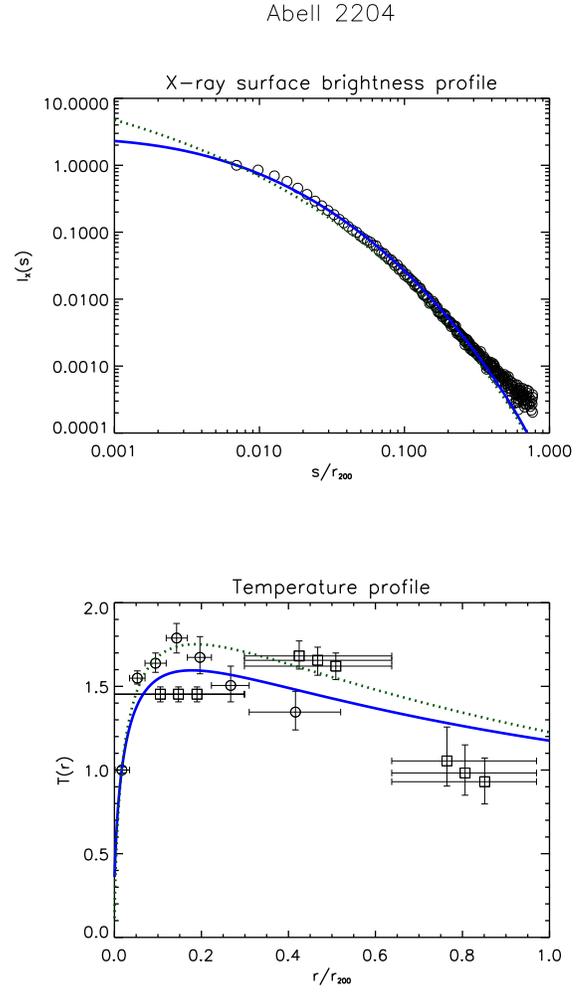}\caption{Same as in
Fig.~5 for the CC cluster A2204.
The solid line represents the Supermodel for a given DM
$\alpha=1.27$-profile with concentration $c= 4$, and ICP
entropy with slope $a= 1.1$ and central value
$\bar{k}_c= 10^{-3}$. The dotted line refers to the
approximation presented in \S~6.1.}
\end{figure}

\begin{figure}
\epsscale{1.2}\plotone{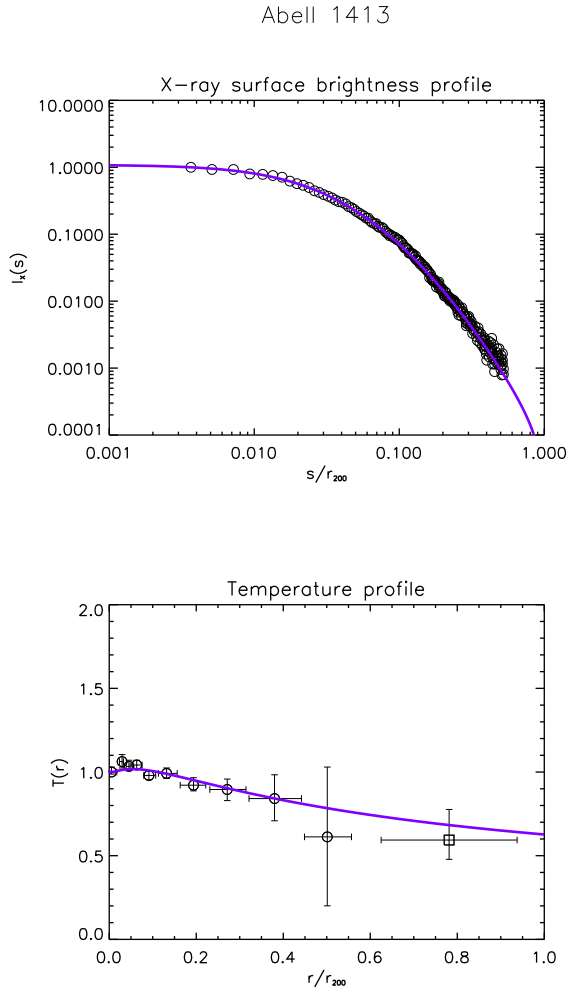}\caption{Same as in
Fig.~5 for the UNC cluster A1413.
The solid line represents the Supermodel for a given DM
$\alpha=1.27$-profile with concentration $c= 5.5$, and
ICP entropy with slope $a= 1.1$ and central value
$\bar{k}_c= 2.5\times 10^{-2}$.}
\end{figure}

\section{The Supermodel for CC and NCC clusters}

The run of $k(r)$ in the handy form Eq.~(17) may be inserted
into the hydrostatic equilibrium Eqs.~(8) to yield the
\emph{Supermodel} for the ICP disposition. Away from the
boundary this links $T(r)$ and $n(r)$ \emph{inversely}; such an
inverse link is particularly clear at the very center, as is
seen from the dependencies $T_c(k_c)$ and $n_c(k_c)$ obtained
from Eqs.~(8) on considering the explicit scalings $T_c \propto
k^{3/5}_c$ and $n_c \propto k_c^{-3/5} $ with the further
factor supplied by the dominant integral that scales as
$k_c^{-0.25}$ (see Appendix A). Thus the overall scaling laws
read
\begin{equation}
T_c \propto k^{0.35}_c~,~~~~~~~~~~~~~~~~~ n_c \propto
k^{-1}_c~;
\end{equation}
these yield $T_c \propto n_c^{-0.35}$, from which we see
clearly how low/high $T_c$ correspond to high/low $n_c$. In
addition, it it is easily perceived, and is seen from Fig.~3,
that the central runs of $n(r)$ are angled or core-like,
depending on $k_c$ being low or high.

There is much more. Fig.~3 show that the Supermodel provides
simultaneous, \emph{accurate} descriptions to both the
\emph{full} profiles of $T(r)$ as directly given by Eq.~(8),
and of the surface brightness in X rays provided on integrating
along the l.o.s. the volume emissivity for optically thin
thermal bremsstrahlung; this reads $2.4\times 10^{-27}\,
\mathcal{L}_X$ erg s$^{-1}$ cm$^{-3}$ with
$\mathcal{L}_X\propto n^2\,T^{1/2}$ (emission lines add for
$k_B T\la 2$ keV, see Sarazin 1988). After Eq.~(8) one has
\begin{equation}
\mathcal{L}_X(\bar{r}) \propto \bar{k}^{-9/10}(\bar{r})\,[1 +
2/5\,b_R\, \int_{\bar{r}}^1{\rm d}\bar{r}'~
\bar{k}^{-3/5}(\bar{r}')\, \bar{v}^2_c(\bar{r}')/\bar{r}']^{7/2}~,
\end{equation}
with the central scaling given by $\mathcal{L}_X (0) \propto
k_c^{-1.8}$. These profiles depend: strongly on the value of
$k_c$ that primarily governs the central pressure and hence the
central density run (see Fig.~3); weakly on $a$ (and $b_R$)
that governs the middle run; mildly on the DM concentration $c$
(see \S~2) that governs the outer decline toward the boundary
values.

We illustrate in Figs.~3 and 4 how straightforwardly the
Supermodel describes various observables for both classes of
Cool Core (CC) and Non Cool Core clusters (NCC) as identified
by Leccardi \& Molendi (2008), and also for their intermediate
class of UNC clusters. In Figs.~5, 6, and 7 we focus on the
specific cases of the clusters A2218, A2204, and A1413 for
which both high-resolution \textsl{XMM-Newton} and preliminary
\textsl{Suzaku} data are available.

From these results is clearly seen how CC clusters are marked
out by the presence of a \emph{peak} of $T(r)$ at $r \approx
0.1 - 0.2 \,R$ (or equivalently by $T_c < T_R$). The condition
for the peak to occur is highlighted on recalling from Eq.~(8)
that
\begin{equation}
T(r) \propto k(r) \, n^{2/3}(r)
\end{equation}
applies, as the ICP counterpart of Eq.~(6) for the DM. Given
that $n(r)$ rises monotonically inward, $T(r)$ will peak and
then decline toward the center when $k^{3/5}(r)$ decreases
strongly toward a low value of $k_c$, as is the case with the
CC clusters. On the other hand, $T(r)$ will rise to a central,
roughly isothermal \emph{plateau} for sufficiently high values
of $k_c$.

From the condition for a maximum to occur in the functional
form of $T(r)$ as given by Eq.~(8), on using $v^2_c (r)$ from
the $\alpha$-profiles with $\alpha = 1.27 - 1.3$ the threshold
value for the peak reads
\begin{equation}
\bar{k}_c \approx 2.5\times 10^{-2} ~;
\end{equation}
with boundary values $k_R \approx 1500-2000$ keV cm$^2$, these
correspond to $k_c \approx 40 - 50$ keV cm$^2$. In closer
detail, the peak looms out (the case of UNC clusters) for
$\bar{k}_c \approx 2.5 \times 10^{-2}$, and stands out (the
case of CC clusters) for $\bar{k}_c\la 10^{-2}$ corresponding
to $k_c \approx 15 - 20$ keV cm$^2$.

We stress that a \emph{finite} $T_c \neq 0$ constitutes a
natural feature of the equilibrium for CC clusters rather than
some peculiarity of cooling flows (see discussion by Peterson
\& Fabian 2006). This holds with realistically small values of
$k_c$; but even if $k_c$ were formally null, $T_c$ would
decline toward the very center with a somewhat flatter powerlaw
$r^{3\,a/5}$ compared to $\sigma^2 \propto r^{3\alpha/5}$. This
behavior is consistent with the nonradiative runs of the
hydrodynamical simulations by Borgani (2007, see his Fig.~1).

A feature typical of the Supermodel is the peak of $T(r)$
closely following the maximum of the DM velocity
\emph{dispersion} $\sigma^2(r)$, not that of the circular
velocity $v^2_c(r)$, see Fig.~3 (middle right panel). As the
former moves considerably downward with masses ranging from
$10^{15}$ to $10^{13}\, M_{\odot}$ (see Fig.~1 and LC09), we
predict the peak of $T(r)$ should also \emph{move} to
progressively lower radii in going from rich to poor clusters
and groups; preliminary data by Nagai et al. (2007) support the
prediction, but robust data require secure \textsl{Chandra}
calibrations.

Finally, the Comptonization parameter for the SZ effect
(Sunyaev \& Zel'dovich 1972) obtains from integrating along a
l.o.s. the volume quantity $\mathcal{Y}\propto p$ in terms of
the (thermalized) electron pressure $p=n\,k_B T$; after Eq.~(8)
this means
\begin{equation}
\mathcal{Y}(\bar{r}) \propto [1 + 2/5\,b_R\,
\int_{\bar{r}}^1{\rm d}\bar{r}'~ \bar{k}^{-3/5}(\bar{r}')\,
\bar{v}^2_c(\bar{r}')/\bar{r}']^{5/2}~,
\end{equation}
to the result plotted in Fig.~4. In the central region this
scales as $\mathcal{Y}(0)\propto \bar{k}_c^{-0.65}$.

\section{Stability of the central conditions}

Having shown how the Supermodel focuses the central conditions,
here we pursue the discussion ending \S~3, and argue that they
are robust against energy losses or additions.

\subsection{From Cool Cores to cooling cores, and back}

Plainly, the CC state produced by the Supermodel differs not
only from a cooling flow but also from a freely cooling core;
in fact, cooling is not included in Eqs.~(7) as they stand.
However, the Supermodel focuses the conditions for enhanced
radiation and fast cooling to set in, namely, low though
\emph{finite} $T_c \propto k^{0.35}_c$ \emph{linked} to high
$n_c$ so as to imply
\begin{equation}
t_{c}\approx 0.3\,\left({k_c\over 15\,\mathrm{keV
cm}^2}\right)^{1.2}~\mathrm{Gyr}~.
\end{equation}
Fig.~3 (bottom left panel) illustrates the run $t_c(r)$ of the
cooling time throughout a CC cluster.

So far as it goes for the Supermodel proper. The sequel of the
story is long accepted in general terms, to imply that such an
enhanced radiation will lead to entropy loss, that in turn will
further lower $T_c$ and increase $n_c$, so shortening $t_{c}$
and opening the way for a classic cooling catastrophe to set in
(White \& Rees 1978; Blanchard et al. 1992). A possible happy
end to the cooling story in clusters has been widely proposed
and discussed (see Binney \& Tabor 1995; Cavaliere et al. 2002;
Voit \& Donahue 2005; Ciotti \& Ostriker 2001, Tucker et al.
2007), to the effect that, before the conditions run away into
a full catastrophe, the ICP condensing around central massive
galaxies and mixing with their ISM is very likely to kindle up
AGN activities by renewing mass accretion onto their
powerhouses, the supermassive BHs lurking and starving at most
galactic centers. Thus recurrent loops are conceivably started
out by cooling that rekindles AGNs, that in turn feed back
energy into the surrounding medium to eventually quench the BH
accretion flow and ultimately yield quasi-steady, widespread
conditions with $k_c$ reset over timescales of some $10^8$ yr
to values around $15$ keV cm$^2$.

\subsection{Saturation in Non Cool Cores}

On the other hand, NCC conditions prevail above the divide at
$k_c\approx 40 - 50$ keV cm$^2$. We understand the fewer NCC
relative to the CC clusters primarily on the basis of strong
AGN inputs, in the tail of the AGN luminosity function $N(L)$
(e.g., Richards et al. 2006); this implies for the output
statistics $\Delta E\, N(\Delta E)\propto L\, N(L)$ a sharp
decline above $\Delta E\approx 5\times 10^{61}$ erg. We also
note that hotter cluster centers tend to impair or prevent the
supersonic condition necessary for driving strong pressurized
blastwaves (see Appendix B), i.e., $\mathcal{M}^2\approx
1-\Delta E/E\ga 3$ corresponding to $\Delta E/2\, k_B T_c\, m
\ga 1$; meanwhile the entropy deposited reads $k_c\approx
\Delta E/(1- \Delta E/2E)^{2/3}$. This will saturate the
effects of multiple inputs, if any, occurring within a cooling
time.

Similar arguments also apply to the possibly larger if
generally rarer energy outputs associated to substantial
mergers; these may be up to $\Delta E \la 10^{64}$ ergs (see
\S~3.1.3), but by the same token the energies effectively
transferred to the ICP in NCC conditions are especially prone
to saturate after the first such event.

\subsection{Concluding on central conditions}

Taking up from \S~3.1.3, we recall that the
cooling-fueling-feedback machinery calls for a correlation
between CC clusters and current central AGNs (the so called
dichotomy, see Voit 2005) that finds support from a
considerable body of observations (e.g., Mittal et al. 2009).
This picture is currently under scrutiny in study cases such as
provided by the poor cluster AWM4; there the considerable value
$k_c\approx 60$ keV cm$^2$ calls for a fueling process
unrelated to cooling, with energetics higher than implied by
the current radio activity. In addition, no signatures or
fossil imprints have yet been found of large energy inputs
caused over the past $10^{-1}$ Gyr by either major AGN
outbursts or mergers (Gastaldello et al. 2008, Giacintucci et
al. 2008). This may constitute as of today one instance of an
exceptional preheating level, standing out as a main component
to $k_c$.

To complete the picture, AGNs as widespread agencies for rising
the central entropies lead to understand the steep decline of
the local $L_X - T$ correlation from clusters to groups, or the
equivalent saturation in groups of the $k - T$ correlation (see
Ponman et al. 2003; Pratt et al. 2009), in terms of comparable
single outbursts in differently sized galaxy systems (Cavaliere
et al. 2002, LCM05). Finally, AGN kinetic plus radiative
outputs also lead to expect for clusters a non monotonic rise
and fall of the relation $L_X -z$, consistent with the current
data (Cavaliere \& Lapi 2008).

In sum, the two main modes for energy injections, namely,
central AGNs and major mergers, provide different levels of
entropy input; whence we expect a bimodal distribution for the
observed number of clusters as a function of the central
entropy level $k_c$. In fact, the two peaks should be remolded
to an actual distance considerably smaller than the factor
$10^2$ separating the two maximal input levels; this is because
the statistics will be eroded at low $k_c$ by fast cooling,
while limited at high $k_c$ by the small number of strong input
events. The expected outcome will be not unlike the findings
recently presented by Voit (2008) and Cavagnolo et al. (2009).

Back to our main course, we conclude that both the CC and the
NCC central conditions envisaged by the Supermodel are made
\emph{robust} by processes additional, but naturally geared to
it.

\section{Limiting models}

While the Supermodel can yield accurate representations of the
ICP state with a moderate amount of formalism, it is
nevertheless worthwhile to have at hands simple limiting models
amenable to prompt analytical computations.

\subsection{Mirror dispersions}

We take up the point made in \S~4 as to $T(r)$ following
$\sigma^2(r)$ for CC clusters, and show in Figs.~3 (middle
right) and 6 (bottom panel) the good performance around the
$T(r)$ peak and shortward (but for the very central range where
$T(r)$ deviates upward to its value $T_c \propto k^{0.35}_c$)
provided by the simple model with the ICP mirroring the DM
dispersion
\begin{equation}
T = \sigma^2/\beta ~.
\end{equation}
The normalizations are included in the constant parameter
$\beta \equiv \mu m_p\, \sigma^2/k_B T$ that in Fig.~6 is fixed
at $0.75$, just the natural value it takes when evaluated at
$R$.

This is similar to the approximation discussed by Cavaliere \&
Fusco-Femiano (1981), and similarly yields the density in the
explicit form
\begin{equation}
\bar{n}(\bar{r}) = \bar{\rho}^{\beta} (\bar{r})\,
\bar{\sigma}^{2 (\beta -1)}(\bar{r})~.
\end{equation}
Here the DM density $\rho(r) $ is provided by the
$\alpha$-profiles of \S~2, to imply
\begin{equation}
n(r) \propto r^{\alpha(\beta -1)}\, \rho ^{5\beta/3 - 2/3}(r)~,
\end{equation}
which goes into the simple form $n(r) \propto \rho(r)$ toward
the center in the range where $\rho(r) \rightarrow r^{3\,
\alpha/5}$ holds but still Eq.~(25) applies.

\subsection{Polytropic $\beta$-models}

For NCC clusters, instead, we make contact with the classic
$\beta$-models discussed by Cavaliere \& Fusco-Femiano (1978).

In the central range the contact obtains on noting these
clusters to be marked by high values of $k_c$ that cause a
nearly flat central run of the entropy after Eq.~(17). It is
now matter of straightforward algebra to see that on taking
$k(r)\approx k_c$ to a zeroth approximation, this may be
extracted from the integral in Eqs.~(8), to yield directly
\begin{equation}
T/T_c= (n/n_c)^{2/3} = 1 + 2\,\beta_c \,
\Delta \phi_{\rm c,r}/5
\end{equation}
corresponding to a polytropic approximation with macroscopic
index $\Gamma \equiv 5/3 + \mathrm{d}\log\, k/\mathrm{d}\log\,n
= 5/3$. To a next approximation $\Gamma$ may be obtained by
carrying further the expansion of the integral, to obtain
\begin{equation}
\Gamma\simeq {5\over 3}\,\left[1-{2\over 5}\, {1-\bar{k}_c\over
\bar{k}_c}\, {1\over \Delta
\phi_{\rm c,r}}\,\int_c^{\bar{r}}{\mathrm{d}\bar{r}'\,
{\mathrm{d}\phi\over \mathrm{d}\bar{r}'}\, \bar{r}'^a}\right]~,
\end{equation}
to be used in the expression
\begin{equation}
T/T_c= (n/n_c)^{\Gamma -1} = 1 + (\Gamma -1)\, \beta_c \,
\Delta \phi_{\rm c,r}/\Gamma ~.
\end{equation}

In the outer regions a similar expansion may be pursued both
for CC and NCC clusters, to yield an effective index
\begin{equation}
\Gamma\simeq 1+{2\over 5}\, {1\over \Delta
\phi_{\rm r,R}}\,\int_{\bar{r}}^1{\mathrm{d}\bar{r}'\,
{\mathrm{d}\phi\over \mathrm{d}\bar{r}'}\, \bar{r}'^{-3\,a/5}}~;
\end{equation}
In fact, Fig.~3 last panel shows that values roughly constant
in the range $\Gamma = 1.1 - 1.2$ apply to the outer regions of
all cluster categories. \footnote{For a quick evaluation of the
index, consider that $\Gamma \equiv 5/3 + d\, \log\, k/d\,
\log\,n = 5/3 - a_R/g_R = b_R/g_R ~.$}

Another approximation of the polytropic type is seen to apply
for $r > r_m$, i.e., to the right of the peak of $\sigma^2$
(see Fig.~1), and may be formally based as follows. Consider
that wherever density and temperature follow (piecewise)
powerlaw runs, the elimination of $r$ provides a link of the
polytropic form $n\, T \propto n^{\Gamma}$; a similar
consideration applies to the DM, leading to define a
corresponding index $\Lambda$. Thus we may write the first and
the third sides of Eq.~(2) in these terms, and equate them
\emph{directly}; then simple algebra provides the explicit
relation
\begin{equation}
{T\over T_m} \approx \beta_m \, {\Lambda\over
\Gamma}\,{\Gamma-1\over \Lambda-1}\, \left({\sigma\over
\sigma_{m}}\right)^2 + {T_R\over T_m} ~.
\end{equation}
This, complementarily to Eq.~(24), shows that the temperature
run $T(r)$ tends to follow the DM dispersion $\sigma^2(r)$
except for the vicinity of the virial boundary where the shock
condition sustains it at the value $T_R$, and for the very
center where a finite if small $k_c$ matters.

Summarizing the trend highlighted by the limiting models, the
passive \emph{mirror} behavior of the ICP with $T\propto
\sigma^2$ prevails unless is offset by energy inputs, as in
fact occurs at the boundary for all clusters, and in the
central region for the NCC clusters.

\section{Discussion and conclusions}

This paper introduces a novel look to the Astrophysics of
galaxy clusters, in terms of both the $\alpha$-profiles for the
initially cold DM and of the Supermodel for the hot ICP.

As for the DM halos, we have taken up from LC09 the
\emph{physical} $\alpha$-profiles. These are based on the Jeans
equilibrium between self-gravity and pressure modulated by the
DM `entropy' run $K(r) \propto r^{\alpha}$. The latter is found
from many recent $N$-body simulations (recalled in \S~1) to
apply with $\alpha$ closely constant within the halo body; we
have semianalytically computed the halo two-stage development
and obtained the narrow range $\alpha\approx 1.27-1.3$ from
poor to rich clusters (see \S~2 for details). The ensuing
$\alpha$-profiles, depending on the \emph{two} key parameters
$\alpha$ and $c$, provide density runs $\rho(r)$ that satisfy
\emph{physical} central and outer boundary conditions at
variance with the empirical NFW formula, and also yield better
fits to detailed data from gravitational lensing in and around
massive clusters (see Lapi \& Cavaliere 2009b).

The ICP, on the other hand, settles to equilibrium within the
gravitational wells associated with the $\alpha$-profiles,
under control from the thermodynamic entropy produced by
boundary and central shocks driven by AGNs or major mergers,
plus a possible preheating basal level (see \S~3). These
\emph{physical} effects may be compounded in the two-parameter
form $\bar{k}(r) = \bar{k}_c + (1-\bar{k}_c)\, \bar{r}^{a}$
with $a$ ranging from $0.8$ to $1.1$. The resulting equilibrium
may be concisely rendered as a trend for the ICP to follow the
DM in the \emph{passive} behavior $T(r) \propto \sigma^2(r)$,
in the radial range \emph{free} from the energy inputs that
steadily produce the \emph{outer} boundary slope and
intermittently refresh the \emph{central} level $\bar{k}_c$.

In detail, our \emph{Supermodel} of \S~4 provides accurate and
extended representations for the runs of ICP temperature and
density and of the related ICP observables (see Fig.~5, 6, and
7 for examples). These validate the assumption of hydrostatic
equilibrium, and closely constrain not only the value of the
concentration $c$ for the DM $\alpha$-profile but also the
\emph{two} ICP parameters $a$ and $\bar{k}_c$. In fact, such
representations hold for either Non Cool Core and Cool Core
clusters, marked out, respectively, by a \emph{monotonic} outer
decline of $T(r)$ from a central plateau at $T_c\ga T_R$, or by
a middle \emph{peak}. In the Supermodel these morphologies are
produced by the central entropy $k_c$ being higher or lower
than a threshold value $k_c \approx 20-50$ keV cm$^2$;
correspondingly, the X-ray brightness features a flat corelike
or a steep central run, based upon the \emph{same} mildly
cusped DM $\alpha$-profile. In a forthcoming paper, we will
present a detailed analysis of an extended sample of NCC and
UNC clusters, with the aims of disentangling the origin of the
central energy inputs, and of evaluating the variance in the
cluster ages through their outer DM concentrations (see \S~2).

We stress that the Supermodel links $T(r)$ and $n(r)$
\emph{inversely} from the bulk toward the central region (see
\S~4); in particular, it yields for the very central values the
scaling $T_c \propto k_c^{0.35}$ and $n_c \propto k_c^{-1}$,
implying $t_{c}\propto k_c^{1.2}$ for the cooling time. The
stability of such values is argued in \S~5. High $T_c$ combines
with flat $n_c$ to produce in NCC clusters central conditions
conducive to \emph{saturation} of $k_c$ toward values around
$10^2$ keV cm$^2$. In CC clusters, instead, low though
\emph{finite} $T_c$ combines with high $n_c$ into a condition
paving the way to fast cooling; this condition is conducive to
triggering intermittent, recurrent loops going through the
stages: cooling, massive BH fueling, AGN energy feedback, that
halts further fueling and activity; these loops make possible
in the long term a quasi-steady state. In the center of NCC
clusters, instead, hot conditions suppress AGN reactivations
owing to lack of dense cool ICP crowded around the central
cluster galaxies; in addition, they tend to saturate the
effective energy coupling from the most powerful AGNs or
mergers by preventing or impairing supersonic conditions
conducive to strong shocks.

Concerning central conditions, we emphasize two points. First,
in CC clusters the Supermodel predicts a \emph{finite}
(non-zero) central $T_c$ with no need for any twist in cooling
flow theories; this rather constitutes a natural condition set
by their equilibria at low $k_c$ and stabilized against cooling
by recurrent AGN activity. Second, in NCC clusters we expect
\emph{saturation} to enforce stability of the higher $k_c$
levels set by the inputs from powerful AGNs and from the
stronger if rarer mergers.

Moving into the middle radial range where energy sources may be
neglected, the `mirror' behavior of Eqs.~(24) and (31) prevails
with $T(r)$ \emph{passively} following $\sigma^2(r)$; a novel
feature emerging from the Supermodel is the very close location
of the two respective peaks. This is because such two
homologous quantities arise from a \emph{parallel} response to
the requirement of withstanding the common gravity for
equilibrium (see \S~6, also Fig.~3). This gravity-induced
behavior is at the root of the remarkable effectiveness of the
simple model $T(r)\propto \sigma^2(r)$, which for the CC
clusters holds well down toward the center (see Fig.~6). As a
consequence, the Supermodel predicts the peak of $T(r)$ to move
toward progressively \emph{smaller} radii in going from rich
clusters to groups.

On approaching the boundary, the run of $T(r)$ again deviates
upward from this passive trend (see \S~3.1 and Appendix B),
with a boundary value $T_R$ sustained by the energy input
associated to infall. Here the passive ICP behavior is again
expected to be broken by the energy transfer due to
electromagnetic interactions and \emph{localized} to a range
$\Delta r\sim \lambda_{pp}\ll r$ (whilst bubbles or shocks
starting from the center smear their energy out to some $10^2$
kpc, implying an effective $\Delta r/r\sim 1$). However,
pinning down the outer deviations requires high sensitivity,
currently achievable only with full use of the \textsl{Suzaku}
capabilities.

\begin{figure}
\epsscale{1.1}\plotone{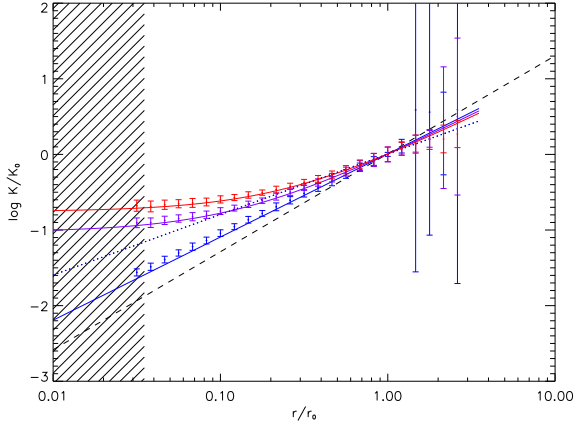}\caption{Reconstruction of the entropy profiles
from joint X-ray and SZ observations (Cavaliere et al. 2005,
see their Eq.~[8]). The dashed line shows the DM entropy
profile with the slope $\alpha=1.3$. Solid lines are the
input ICP entropy profiles described by Eq.~(17), for our
standard values $\bar{k}_c=5\times 10^{-2}$ (red),
$\bar{k}_c=2.5\times 10^{-2}$ (cyan), and $\bar{k}_c=0$ (blue);
the error bars illustrate the reliability of the reconstruction
from joint mock observations in a rich cluster at $1$ Gpc of
the X-ray brightness with resolution of $1''$ and
sensitivities $\Delta I_X = 2\times 10^{-4}\, I_{Xc}$, and of
the SZ effect with resolution of $10''$ and sensitivities
$\Delta I_y = 10^{-2}\, I_{yc}$. The blue dotted line
represents an alternative ICP entropy profile with $a=0.8$
(and $\bar{k}_c=0$) as we expect for ICP preheated at $1/2$ keV
per particle (see \S~3.1.1); similar shapes have been observed by
Sun et al. (2009) in many poor clusters, and by Lemze et al.
(2008) in A1689.}
\end{figure}

Next we highlight an unexpected \emph{connection} specifically
emerging from the Supermodel. X-ray observations of clusters
yield information concerning the concentration $c$ through the
values of the outer entropy slope $a$, and more directly from
detailed fits to the surface brightness data; in fact, we
expect $a$ to be \emph{lower} and the density profile to be
steeper for early clusters with \emph{higher} concentrations
$c$ (see \S~3, in particular below Eqs.~[13]). This specific
prediction may be tested through extended simulations covering
high-$c$ halos and nonadiabatic processes as discussed, e.g.,
by Borgani (2007). In parallel, high concentrations are keenly
pinpointed by gravitational lensing observations (Lapi \&
Cavaliere 2009b and references therein). This opens the way to
the use of existing X-ray data as convenient pointers to
targets for time-expensive gravitational lensing observations.

To conclude, we turn to contrast the ICP and the DM behaviors.
Note that the two density runs
$\rho(r)\propto[\sigma^2(r)/K(r)]^{3/2}$ and $n(r)\propto [k_B
T(r)/k(r)]^{3/2}$ will \emph{differ} even where $T\propto
\sigma^2$ applies, to the extent that the DM and ICP entropy
runs differ. This brings us to directly compare these two
governing entropies.

In a nutshell, their \emph{common} features stem from smooth,
slow gravitational mass infall onto the outskirts, while their
detailed runs both in the outer and in the central range
reflect their different sensitivity to other energy inputs.
Quantitatively, both the underlying key parameters $b_R$ and
$\kappa_{\rm crit}$ are amenable to \emph{conversion} of infall
kinetic energy. They take on very close values $b_R \approx
2.65-2.55$ (see \S~3) and $\kappa_{\rm crit}\approx 2.6-2.5$
(see \S~2) at their corresponding fiducial points $r\approx R$
or $r = r_p$; for increasing $\alpha$ or concentration $c$,
they decrease together since both depend on $1/\Delta\phi$ in
terms of the relevant potential drops from the turning point to
$R$ or $r_p$ (see Eq.~[10] for the ICP and Eq.~[13] in LC09 for
the DM).

On the other hand, the \emph{differing} features stem from
local vs. non-local character of the energy conversion. The
collisionless DM particles fall from the cluster surroundings
well into the body, where their kinetic energy is
\emph{non-locally} and progressively randomized, and spreads
out entropy by orbit superposition and stratification with
widely distributed apocenters (see LC09 and references
therein). Correspondingly, in the DM halos $K(r)\equiv
\sigma^2/\rho^{2/3}$ starts out in the outskirts with uniform
values $\alpha \sim 1.1$ in all clusters, to \emph{steepen}
toward the body to universal values $1.27-1.3$ in a gently
convex shape\footnote{Note that our slope of $K \equiv
\sigma^2/\rho^{2/3} $ defined in terms of the $1$-D velocity
dispersion is consistent with that in terms of the $3$-D
dispersion, e.g. Faltenbacher et al. (2007); in detail, Dehnen
\& McLaughlin (2005) and Ascasibar \& Gottl\"{o}ber (2008) find
the latter to be somewhat flatter than the former in the outer
body where radial anisotropies tend to prevail.}.

Meanwhile, in the ICP $k(r)\equiv k_B T/n^{2/3}$ starts out at
the boundary $r\approx R$ from somewhat lower average values
$a\approx 1.1-1.2$, but with a considerable \emph{variance}
when large preheating and high DM concentrations are included,
see \S~3.1.1 and evidence referred to therein. The slope, if
anything, \emph{flattens} out toward an effective value
$a\approx 0$ in the presence of central energy inputs.

Beyond details, this concave (vs. convex) shape of $a$ toward
the center, together with the sensitivity of its bondary values
to outer potential and preheating constitute features specific
to the ICP. Thus we conclude that basically \textit{similar}
gravitational processes in DM and ICP (randomization of bulk
kinetic energy, but on different scales) with the addition of
the ICP collisional sensitivity to other energy inputs, concur
to produce \emph{dissimilar} shapes for $K(r)$ and $k(r)$.

Model independently, we propose two observational tests
addressed at \emph{directly} probing in clusters the two
underlying entropies. The run $K(r)$ of the DM entropy can be
derived from probing the $\alpha$-profile by gravitational
lensing observations as recalled in \S~2. The run $k(r)$ of the
ICP entropy can be reconstructed as proposed by Cavaliere et
al. (2005) and illustrated in Fig.~8 starting from the relation
\begin{equation}
k(r) = \mathcal{Y}^{14/9}(r)\,\mathcal{L}_X^{-10/9}(r)~,
\end{equation}
that joins deconvolved observations of X-ray brightness and SZ
effect irrespective of any modeling or assumption on
hydrostatic equilibrium and of redshift information (Cavaliere
et al. 1977). Such studies may be particularly useful in the
ongoing search for early clusters (see Andreon et al. 2009).

\begin{acknowledgements}
We thank an anonymous referee for keen comments and helpful
suggestions. We have benefited from various exchanges with A.
Biviano, S. Borgani, A. Diaferio, M. Norman, Y. Rephaeli, and
P. Rosati. Work partially supported by ASI and INAF. AL thanks
INAF-OATS for kind hospitality.
\end{acknowledgements}

\begin{appendix}

\section{The integral in Eq.~(8)}

Here we provide an effective approximation to the integral
$I(r)$ appearing on the right-hand side of Eq.~(8), in terms of
the following analytical expression
\begin{equation}
I(\bar{r})\equiv \int^1_{\bar{r}}\, {\mathrm{d}\bar{r}'\over
\bar{r}'}\, \bar{v}^2_c(\bar{r}')\,
\bar{k}^{-3/5}(\bar{r}')
\simeq A_0\,\exp(-A_1\,\bar{r}^{~A_2})~;
\end{equation}
the fitting parameters $A_0$, $A_1$, and $A_2$ depend weakly on
$\bar{k}_c$, as specified in Table~A1. The dotted lines in
Fig.~2 (top panel) illustrate the effectiveness of such an
approximation for our three standard values of $\bar{k}_c$.
Note that on approaching the center the integral is numerically
found to scale with central entropy $k_c$ like $I_c \propto
k_c^{-1/4}$, see Fig.~2 (bottom panel). This result is used in
\S~4 of the main text.

\begin{deluxetable}{cllllllllllll}
\tabletypesize{} \tablecaption{Fit parameters of Eq.~(A1)}
\tablewidth{0pt} \tablehead{\colhead{} & \colhead{} &
\multicolumn{3}{c}{$c=3.5$} & \colhead{} &
\multicolumn{3}{c}{$c=4.5$} & \colhead{} &
\multicolumn{3}{c}{$c=5.5$}\\
\cline{3-5} \cline{7-9} \cline{11-13} \\
\colhead{$\bar{k}_c$} & \colhead{} & \colhead{$A_0$} &
\colhead{$A_1$} & \colhead{$A_2$} & \colhead{} &
\colhead{$A_0$} & \colhead{$A_1$} & \colhead{$A_2$} &
\colhead{} & \colhead{$A_0$} & \colhead{$A_1$} &
\colhead{$A_2$}} \startdata 0& & 17.77& 2.86& 0.49& & 22.82
&2.72& 0.47& & 27.59&2.64& 0.47\\
10$^{-4}$& & 17.47& 2.86& 0.49& & 22.33& 2.71& 0.48& & 26.88& 2.63& 0.48\\
10$^{-3}$& & 15.93& 2.83& 0.54& & 19.99& 2.66& 0.54& & 23.72& 2.56& 0.53\\
$5\times 10^{-3}$& & 13.42& 2.73& 0.62& & 16.53& 2.51& 0.61& &
19.34& 2.38& 0.61\\
10$^{-2}$& & 12.04& 2.63& 0.66& & 14.70& 2.39& 0.66& & 17.09& 2.25& 0.65\\
$2.5\times 10^{-2}$& & 10.06& 2.44& 0.72& & 12.14& 2.18& 0.71& & 13.99& 2.02& 0.71\\
$5\times 10^{-2}$& & 8.53& 2.26& 0.77& & 10.19& 1.99& 0.75& & 11.66& 1.83& 0.74\\
$7.5\times 10^{-2}$& & 7.64& 2.14& 0.79& & 9.08& 1.87& 0.77& & 10.34& 1.71& 0.76\\
$10^{-1}$& & 7.03& 2.05& 0.81& & 8.31& 1.78& 0.79& & 9.44& 1.62& 0.77\\
\enddata
\tablecomments{In computing the integral Eq.~(8) we have used
the DM $\alpha$-profiles with $\alpha=1.27$. The above fitting
coefficients are given for three values of the concentration
parameter; other values may be derived by standard
interpolation techniques.}
\end{deluxetable}

We plan to provide elsewhere a set of fitting formulae for
$\rho(r)$ and $v_c^2(r)$ leading to an analytic expression for
$I(r)$, as tools enabling direct data analysis and extensive
precision fits.

\section{Entropy production in shocks}

While cooling may condense out the colder fractions of the ICP
and indirectly raise the average entropy of the rest, it is
generally agreed (see Cavaliere et al. 2002; Voit 2005) that
substantial entropy production requires shockwaves. Here we
recall from LCM05 the temperature, density and entropy jumps
produced across a shock transitional layer.

Conservation of mass, energy and total stress across the latter
lead to the classic Rankine-Hugoniot temperature jump
\begin{equation}
{T_2 \over T_1} = {5 \over 16}\, \tilde{\mathcal{M}}^2+{7\over
8}-{3\over 16}\, {1\over
\tilde{\mathcal{M}}^2}~.
\end{equation}
The subscripts $1$ and $2$ denote the pre- and post-shock
quantities, and $\tilde{\mathcal{M}}\equiv (3\,\mu m_p\,
\tilde{v}_1^2/5\,k_B T_1)^{1/2}$ the shock Mach number; the
quantities with tildes refer to the \emph{shock} reference
frame, which is convenient in the case of internal shocks
driven, e.g., by AGNs.

On the other hand, in the case of accretion shocks it is more
convenient to work in terms of the infall velocity $v_1$ and of
the related Mach number $\mathcal{M}\equiv (3\,\mu m_p\,
v_1^2/5\,k_B T_1)^{1/2}$ measured in the \emph{center of mass}
frame. Assuming the downstream kinetic energy to be small, one
finds the temperature jump in the form
\begin{equation}
{T_2\over T_1} = 1 + {4\over 9}\,\mathcal{M}^2\,\left[{1\over
4}+\sqrt{1+{9\over 4}\,{1\over \mathcal{M}^2}}\right]~.
\end{equation}

In either reference frame the density jump in terms of the pre-
and post-shock temperatures reads
\begin{equation}
{n_2\over n_1} = 2\, \left( 1-{T_1 \over T_2}\right) + \sqrt{4\, \left( 1-{T_1
\over T_2} \right)^2 +{T_1 \over T_2}}~.
\end{equation}
The entropy jump $K_2/K_1 = (T_2/T_1)/(n_2/n_1)^{2/3}$ may be
easily composed from the relations above.

Handy expressions apply to \emph{strong} shocks, when the above
expressions reduce to
\begin{equation}
{T_2\over T_1}\simeq {3\over 16}\,{\mu m_p\, \tilde{v}_1^2\over
k_B T_1}+{7\over 8}\simeq {1\over 3}\,{\mu m_p\, v_1^2\over k_B
T_1}+{3\over 2}~,~~~~~~{n_2\over n_1}\simeq 4\,\left(1-{15\over
16}\,{T_1\over T_2}\right)~;
\end{equation}
these approximations actually apply to high/intermediate
$\tilde{\mathcal{M}}^2\ga 3$; the corresponding entropy jumps
read
\begin{equation}
{K_2\over K_1}\simeq {3\over 16}\,{1\over 4^{2/3}}\,{\mu m_p\,
\tilde{v}_1^2\over k_B T_1}+{3\over 2}\,{1\over 4^{2/3}}\simeq
{1\over 3}\,{1\over 4^{2/3}}\,{\mu m_p\, v_1^2\over k_B
T_1}+{17\over 8}\,{1\over 4^{2/3}}~.
\end{equation}
In last relation, the second term on the r.h.s. expresses the
contributions of the advected external entropy $0.84\, K_1$,
and is relevant when $k_B T_1 \ga 0.16\, m_p v^2_1$, that is,
when relatively strong preheating affects the gas infalling
into a poor cluster or a group.

When a central energy pulse is discharged into the equilibrium
ICP with a density gradient, e.g. $n(r) \propto r^{-2}$, a
blast wave is sent out, that is, a non-linear perturbed flow
that terminates into a leading shock and comprises bulk kinetic
energy up to matching the thermal one (see Sedov 1959). When
the pulse is short compared with the transit time and the
effects of gravity and upstream pressure are neglected, the
Mach number decreases radially as $\mathcal{M}(r) \propto
r^{-1/2}$. When the pulse is sustained during the transit time,
also $\mathcal{M}(r)$ will be. Even with gravity and upstream
pressure considered, this is found to hold in the simple case
of a pulse constant over the transit time through the region
where $n(r)\propto r^{-2}$ applies, and to hold also for other
combinations of equilibrium gradients and pulse shapes (LCM05).
For longer times, $\mathcal{M}(r)$ declines and the blast
dissipates its kinetic energy into the ICP.

In such cases the Mach number depends on the energy input
$\Delta E$ relative to the ICP total energy $E$ in the affected
volume, simply as $\mathcal{M}^2\approx 1+\Delta E/E$; the
condition for a strong shock $\mathcal{M}^2\ga 3$ clearly
translates into $\Delta E/2\,m\,k_B T\ga 1$. It turns out that
the ICP may be (partly) evacuated from the central region to a
residual average density $n (1- \Delta n/n) \propto 1-\Delta
E/2\,E$, with an associated entropy $k_c \propto \Delta E/(1-
\Delta E/2\,E)^{2/3}$.

\end{appendix}

\end{document}